\documentclass[conference]{IEEEtran}
\IEEEoverridecommandlockouts
\usepackage{cite}
\usepackage{amsmath,amssymb,amsfonts}
\usepackage{algorithmic}
\usepackage{graphicx}
\usepackage{textcomp}
\usepackage{xcolor}

\usepackage{blkarray}
\usepackage{amsmath}
\usepackage{kbordermatrix}
\usepackage{dsfont}
\usepackage{amsthm}
\usepackage{lipsum}
\usepackage{graphicx}

\usepackage[utf8]{inputenc}
\usepackage[english]{babel}
\usepackage{listings}
\usepackage{fancyvrb}
\usepackage{comment}

\def\BibTeX{{\rm B\kern-.05em{\sc i\kern-.025em b}\kern-.08em
    T\kern-.1667em\lower.7ex\hbox{E}\kern-.125emX}}
\begin{document}

\title{\begin{scriptsize}\textcolor{blue}{\emph{This work has been submitted to the IEEE for possible publication. Copyright may be transferred without notice, after which this version may no longer be accessible.}} \end{scriptsize}\\
A Data Augmented Bayesian Network for Node Failure Prediction in Optical Networks
\thanks{This research is funded by \emph{Tejas Networks}, Bangalore, India}
}

\author{\IEEEauthorblockN{Dibakar Das}
\IEEEauthorblockA{
\textit{IIIT Bangalore}\\
Bangalore, India \\
dibakard@ieee.org}
\and
\IEEEauthorblockN{Mohammad Fahad Imteyaz}
\IEEEauthorblockA{
\textit{Tejas Networks}\\
Bangalore, India \\
imteyaz@tejasnetworks.com}
\and
\IEEEauthorblockN{Jyotsna Bapat}
\IEEEauthorblockA{
\textit{IIIT Bangalore}\\
Bangalore, India \\
jbapat@iiitb.ac.in}
\and
\IEEEauthorblockN{Debabrata Das}
\IEEEauthorblockA{
\textit{IIIT Bangalore}\\
Bangalore, India\\
ddas@iiitb.ac.in}
}

\maketitle

\begin{abstract}
Failures in optical network backbone can cause significant interruption in internet data traffic. Hence, it is very important to reduce such network outages. Prediction of such failures would be a step forward to avoid such disruption of internet services for users as well as operators. Several research proposals are available in the literature which are applications of data science and machine learning techniques. Most of the techniques rely on significant amount of real time data collection. Network devices are assumed to be equipped to collect data and these are then analysed by different algorithms to predict failures. Every network element which is already deployed in the field may not have these data gathering or analysis techniques designed into them initially. However, such mechanisms become necessary later when they are already deployed in the field. This paper proposes a Bayesian network based failure prediction of network nodes, e.g., routers etc., using very basic information from the log files of the devices and applying power law based data augmentation to complement for scarce real time information. Numerical results show that network node failure prediction can be performed with high accuracy using the proposed mechanism.
\end{abstract}


\section{Introduction}
Today's digital world depend primarily on internet. Internet backbone network carries bulk of the data traffic from different users, such as, individuals, Internet of Things (IoT) devices, edges devices, computers and cloud. Backbone networks primarily use optical communications due to their high bandwidth and low bit error rates. These networks comprise of huge number of nodes, e.g., routers, etc., which carry data from one part of the world to the other. A failure in any of these nodes can lead to major disruption in internet services leading to losses in business and other activities. Hence, for reliable internet services it is essential to prevent failures proactively in backbone networks using intelligent mechanisms.

There are several approaches for failure prediction in optical networks. A gaussian classifier based approach to detect single link failures has been proposed in \cite{cite_gaussian_classifier_heuristic_single_failure}. Authors applied heuristics to shortlist the probable failed links and then the gaussian classifier is applied to identify the failed link. \cite{cite_svn_time_series_net_node} proposes support vector machine along with double exponential smoothing approach to predict optical network equipment failure. \cite{cite_supervised_wsn} describes a method for prediction of link quality estimate in wireless sensor networks using online and offline supervised learning. A comparison of three data mining approaches, K-Means, Fuzzy C-Means, and Expectation Maximization, to detect abnormal behaviour in networks is proposed in \cite{cite_clustering_network_traffic_faults}. Using Bayesian networks, \cite{cite_cellular_networks_bn} derives a mechanism to predict failures in cellular networks.

Most of the proposals mentioned above are data intensive. They rely on collecting real time data from various monitors in the network and then analyze the data to predict failures. For deployed systems in the field, such prediction mechanism may not be built into the initial design. However, subsequently, a need for failure prediction arises. In such a scenario, non-availability of relevant data is a major hindrance. Changes to the deployed system like introducing new probes to collect data are highly risky. Hence, applying convention data intensive techniques are not possible. Non-intrusive failure prediction techniques have to be developed with very little information available (quantitative or qualitative) without disturbing the deployed network. This paper proposes such a technique using Bayesian Networks (BN) as explained below. In \cite{cite_comsnets_dag_nonintrusive}, we described an architecture for non-intrusive fault prediction in network nodes. It applies an ad-hoc node failure prediction mechanism as an initial solution. This paper extends and generalizes the network node failure prediction mechanism in \cite{cite_comsnets_dag_nonintrusive} applying formal approach of data augmented BN.

Network nodes are equipped with log files which are used by the developers to debug problems. Observing the logs of past failures, patterns emerge on the sequence of events leading to a failure. These events can be represented as nodes in a Directed Acyclic Graph (DAG). This DAG can used as BN based failure prediction mechanism. Bayesian networks need conditional probabilities of a node (event) given its parents in the DAG for prediction. As already mentioned above, statistics on events and failures are not readily available in deployed network nodes. However, qualitative information on how frequently or infrequently a failure occurs can be acquired from the developers. Using this information, data augmentation is applied to generate the conditional probabilities assuming power law distribution for failure occurrences. The BN uses these probabilities and predicts failures as events occur in real-time. Numerical results show, even with scarce data available from logs retrieved from the deployed network nodes, fairly accurate failure prediction is possible.

Objectives behind this BN based approach are as follows.
\begin{itemize}
\item Construct a quick solution to meet time to market requirements
\item Construct a non-intrusive prediction mechanism devoid of any changes in the deployed network
\item Effectively use information from the logs and qualitative information on frequency of occurrence of failures from the developers
\item Failure prediction mechanism should evolve over time
\end{itemize}

The remaining of this paper is organized as follows. Section \ref{section_model} describes the proposed idea and the system model. The results obtained applying the proposed idea are presented in section \ref{section_results}. Section \ref{section_conclusion} concludes this work with some future directions.
\section{System Model}\label{section_model}
As already mentioned, the statistical information about the occurrence of events and failures at network nodes is not readily known, since the deployed systems are not equipped with necessary mechanisms to collect such data by initial design. Mining all the historical logs to extract statistical information mentioned above can be a extremely time consuming approach and may not meet time to market requirements. The only information extracted from the logs is the sequence of events leading to failures with the help of the developers.  Also, qualitative information on which errors occur more frequently than others can be known from the experience of the developers.

\begin{figure}[ht]
\centering
\includegraphics[width=\columnwidth]{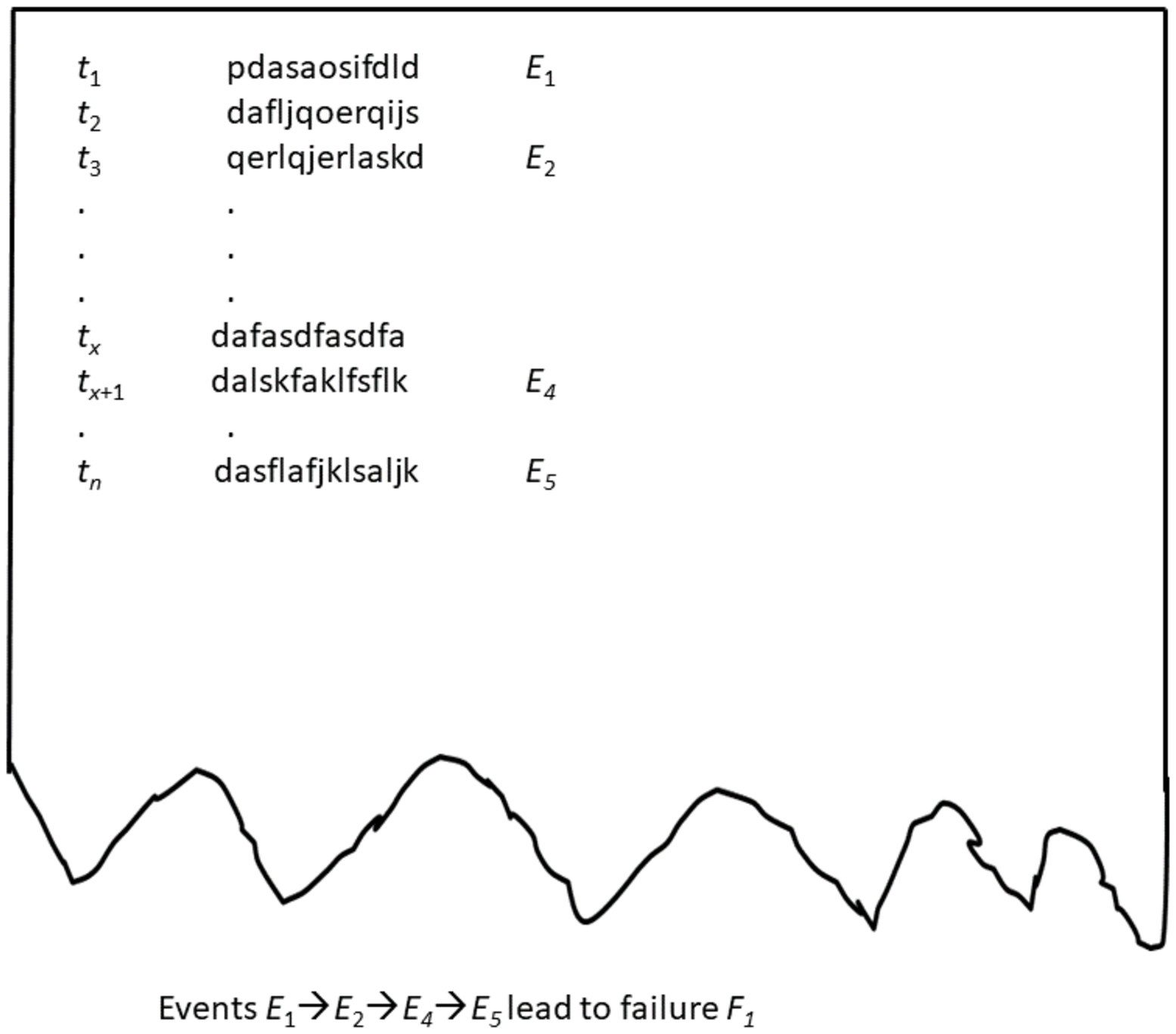}
\caption{Example log file with events}
\label{fig_event_logfile_template_inkscape}
\end{figure}

\begin{figure}[ht]
\centering
\includegraphics[width=\columnwidth]{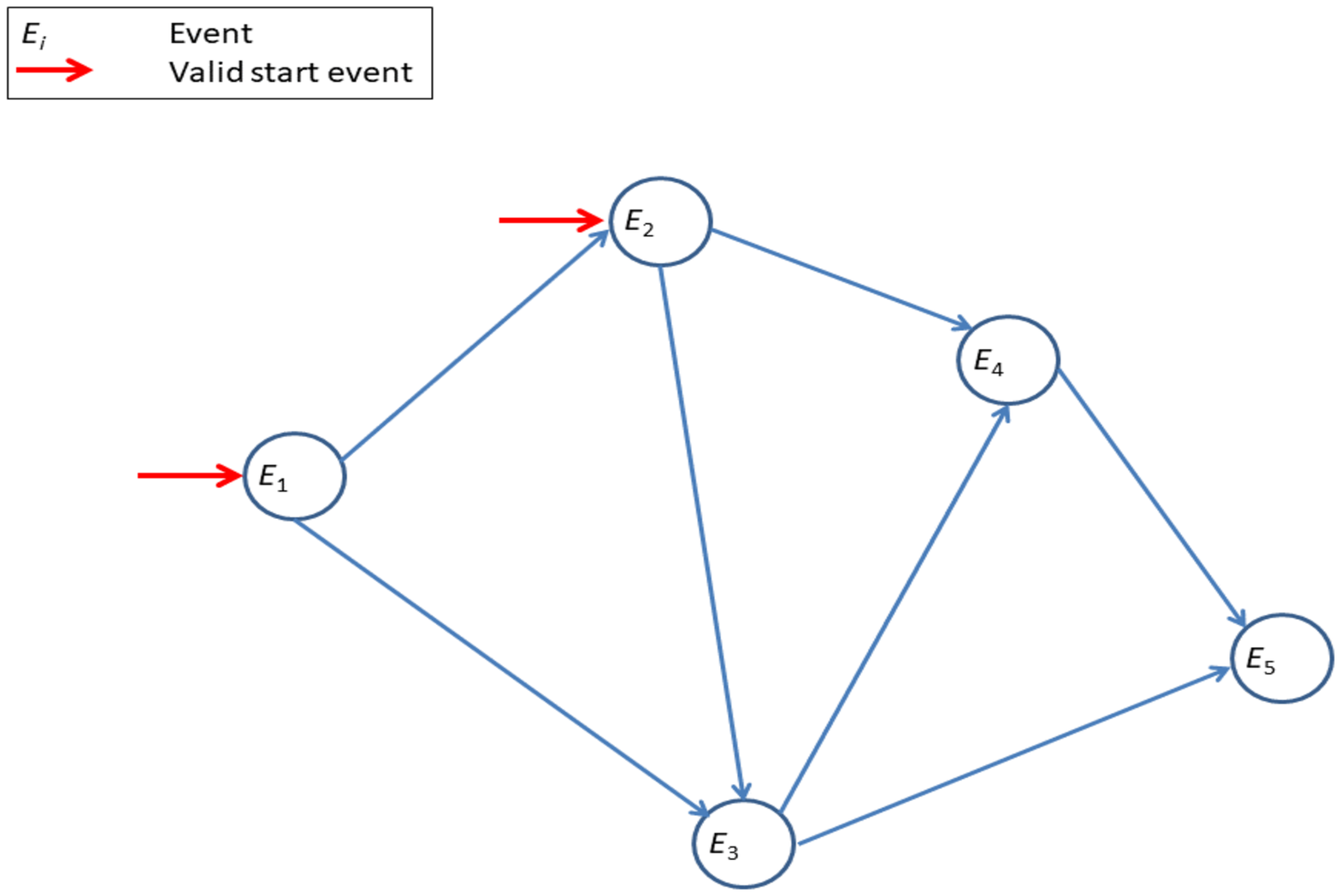}
\caption{DAG constructed using (\ref{eqn_event_failure_table})}
\label{fig_DAG_for_BN_inkscape}
\end{figure}

An example log file is shown in Fig. \ref{fig_event_logfile_template_inkscape}. The first column contains the time at the which the corresponding text (second column) is logged and associated values of system parameters, e.g., clock drift, Optical Signal To Noise Ratio (OSNR), etc. Based on analysis of the developers some of the texts can be designated as events shown in third column of Fig. \ref{fig_event_logfile_template_inkscape}. There can be several events, such as, clock drift exceeding certain threshold, temperature rising above a certain value, OSNR exceeding lower threshold, or a node not receiving signal from its peer. Once the failures and their corresponding events are designed from the logs, they are presented in form of a matrix as shown in (\ref{eqn_event_failure_table}) for 5 failures. Each row in the matrix represents a sequence of events leading to a failure. A value 1 means that the corresponding event has to happen for that particular failure. For example, event $E_2$ has to happen for failures $F_1$, $F_2$ and $F_3$, not for $F_4$ and $F_5$. Subsequently, a DAG comprising all the events can be constructed (Fig. \ref{fig_DAG_for_BN_inkscape}) which forms the BN. For example, event $E_1$ $\rightarrow$ $E_2$ $\rightarrow$ $E_3$ $\rightarrow$ $E_5$ have to occur in sequence for failure $F_2$. Note that $E_1$ and $E_2$ (marked in red) are the valid start states of event sequences leading to  failures. By (\ref{eqn_event_failure_table}), $F_1$, $F_2$, $F_4$ and $F_5$ start with $E_1$, and $F_3$ starts with $E_2$.
\begin{equation}
E_{5,5} = \kbordermatrix{
    & E_1 & E_2 & E_3 & E_4 & E_5\\
    F_1 & 1 & 1 & 0 & 1 & 1\\
    F_2 & 1 & 1 & 1 & 0 & 1\\
    F_3 & 0 & 1 & 0 & 1 & 1\\
    F_4 & 1 & 0 & 1 & 1 & 0\\
    F_5 & 1 & 0 & 1 & 0 & 1\\
  }
  \label{eqn_event_failure_table}
\end{equation}
\subsection{Generation of statistics for events and failures}\label{section_power_law_statistics}
To apply BN for failure prediction, statistics of occurrence of events and their failures are necessary to calculate the conditional probabilities. However, as already mentioned such statistics are not readily available. For this purpose, two available information are used. Firstly, events are extracted from old logs as explained above. Secondly, developers can provide the information on which failures occur more frequently than others. Based on this information, a probability distribution can be assumed to artificially create statistics of the events and their corresponding failures. Since, there is non-zero chance of any failure a scale free probability distribution can be assumed. For this purpose, a power law probability distribution is assumed in (\ref{eqn_power_law}) for occurrence of $N$ failures, though other scale free functions can also be considered (in future).
\begin{equation}\label{eqn_power_law}
p(x) = ax^{-k}
\end{equation}
where $a$ is constant and failure $F_x$ occurs with probability $p(x)$, $x = 1, 2,.., N$ and $k \ge 2$. Also, it is assumed that $p(i) > p(j)$ for $i < j$ and $i, j \in \{1, 2,.., N\}$. Value of $a$ can be adjusted so that,
\begin{equation}\label{eqn_total_prob}
\sum_{x = 1}^N p(x) \approx 1
\end{equation}

Based on the probability distribution function, statistics of each of the failures can be calculated as follows using (\ref{eqn_stats_of_failures}).
\begin{equation}\label{eqn_stats_of_failures}
C_{F_x} = \lfloor S \times p(x)\rfloor
\end{equation}
where $C_{F_x}$ is the number of occurrences of $F_x$ in the population of $S$ failures.

Once the number of occurrence of each failure is estimated with (\ref{eqn_stats_of_failures}), the statistics of the events can also be found out using (\ref{eqn_event_failure_table}). Using all these augmented data and information, a BN can be constructed with all the conditional probabilities.

\subsection{Application of Bayesian Networks}
Application of BN is explained with the following scenerio. Lets evaluate the probabilities of occurrences of $E_4$ and $E_5$ given $E_1 = 1$ (Fig. \ref{fig_DAG_for_BN_inkscape}) as shown in (\ref{eqn_example_predict_0}). Note that there are 4 possible combinations for $E_4$ and $E_5$.
\begin{equation}\label{eqn_example_predict_0}
Pr(E_5, E_4 | E_1 = 1) = \frac{Pr(E_5, E_4)}{Pr(E_1 = 1)}
\end{equation}
$E_4$ and $E_5$  depend on other predecessors (Fig. \ref{fig_DAG_for_BN_inkscape}), so the above equation has to be expanded as follows.

\begin{multline}\label{eqn_example_predict_1}
\Rightarrow Pr(E_5, E_4 | E_1 = 1) = \frac{1}{Pr(E_1 = 1)} \\
                        \times \sum_{E_3}\sum_{E_2}Pr(E_5, E_4, E_3, E_2)
\end{multline}

\begin{multline}\label{eqn_example_predict_2}
\Rightarrow Pr(E_5, E_4 | E_1 = 1) = \frac{1}{Pr(E_1 = 1)} \\
                        \times \sum_{E_3}\sum_{E_2}Pr(E_5 |E4, E3) \\
                        \times Pr(E_4 |E3, E2) \\
                        \times Pr(E_3 | E2, E1) \\
                        \times  Pr(E_2 |E_1)
\end{multline}

\begin{multline}\label{eqn_example_predict_3}
\Rightarrow Pr(E_5, E_4 | E_1 = 1) = \frac{1}{Pr(E_1 = 1)} \\
                        \times \Big\{\sum_{E_3}Pr(E_5 |E4, E3) \times \Big(\sum_{E_2}\\
                        \times Pr(E_4 |E3, E2) \\
                        \times Pr(E_3 | E2, E1) \\
                        \times  Pr(E_2 |E_1)\Big)\Big\}
\end{multline}
Thus, $Pr(E_5, E_4 | E_1 = 1)$ is expressed as conditional probabilities of occurrences of its predecessors in Fig. \ref{fig_DAG_for_BN_inkscape}. These derivations have to be repeated for each combination of events. It is evident, when the BN is large, these derivations can be extremely tedious and cumbersome.
\subsection{Failure Prediction}
Once the BN is constructed as explained above, failure prediction is performed based on the events happening in real time, extracted from network node logs and traversing the BN. A remote machine, running the proposed BN failure prediction model, transfers the real time logs from the network nodes using remote copy, etc., parses the logs for events, using the architecture proposed in \cite{cite_comsnets_dag_nonintrusive}.

\section{Results and Discussion}\label{section_results}
This section presents the results obtained using the system model in section \ref{section_model}. The model is implemented in \verb|python| using  \verb|pgmpy| library \cite{cite_pgmpy_site}. The first step is the generation of statistics of occurrence of failures. Using the generated statistics, the failures are predicted using BN. Calculating the conditional probabilities given all its predecessors of a event in the BN manually for equations such as (\ref{eqn_example_predict_3}) can be extremely cumbersome, tedious and error-prone when the network is large (which is expected to be in future). Hence, using a tool such as \verb|pgmpy| can be extremely beneficial to reliably calculate the probabilities.
\subsection{Generation of failure statistics}\label{section_results_generate_samples}
\begin{figure}[ht]
\centering
\includegraphics[width=\columnwidth]{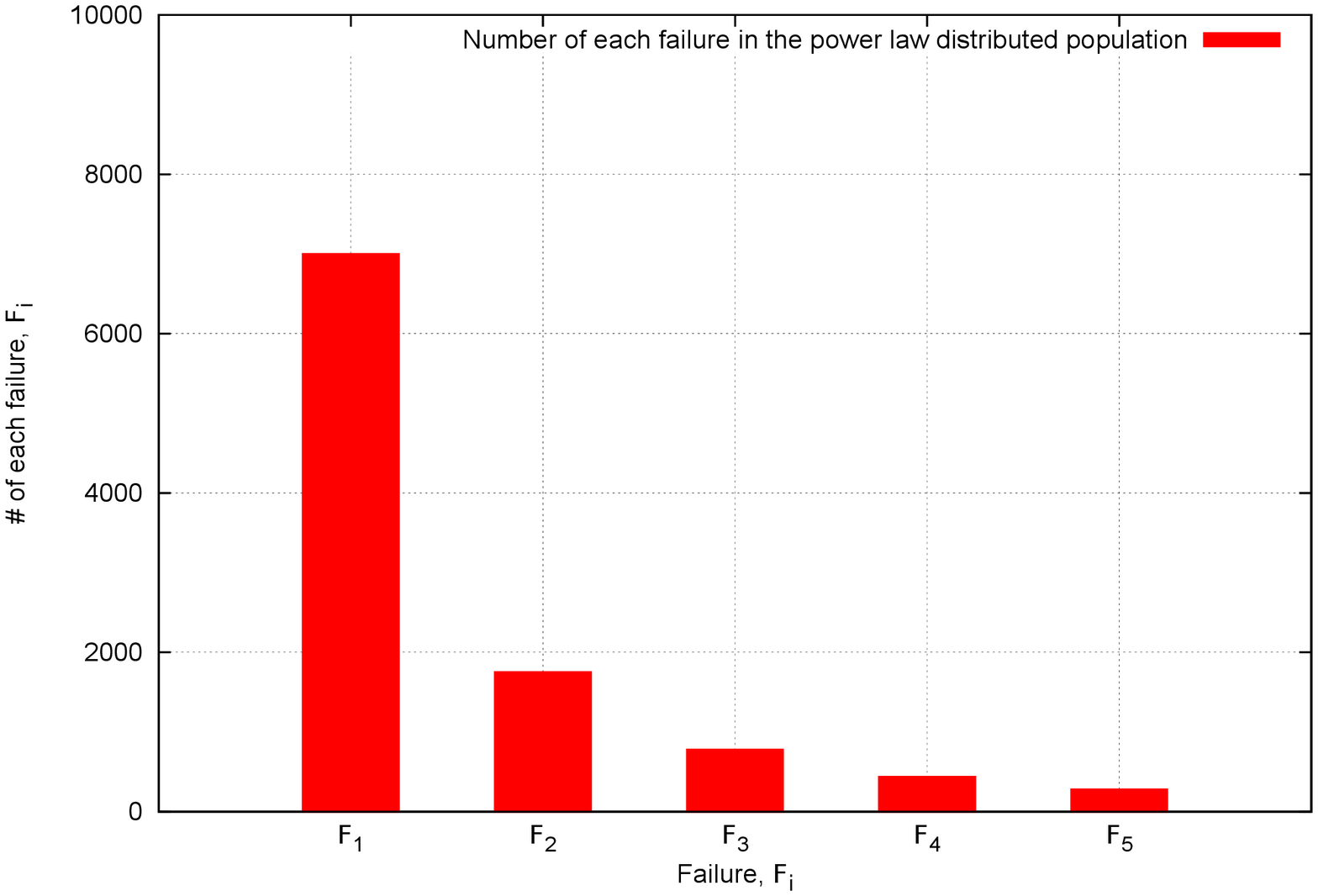}
\caption{Number of each failure $F_i$ in the power law distributed population}
\label{fig_plot_number_of_each_event_inkscape}
\end{figure}

\begin{figure}[ht]
\centering
\includegraphics[width=\columnwidth]{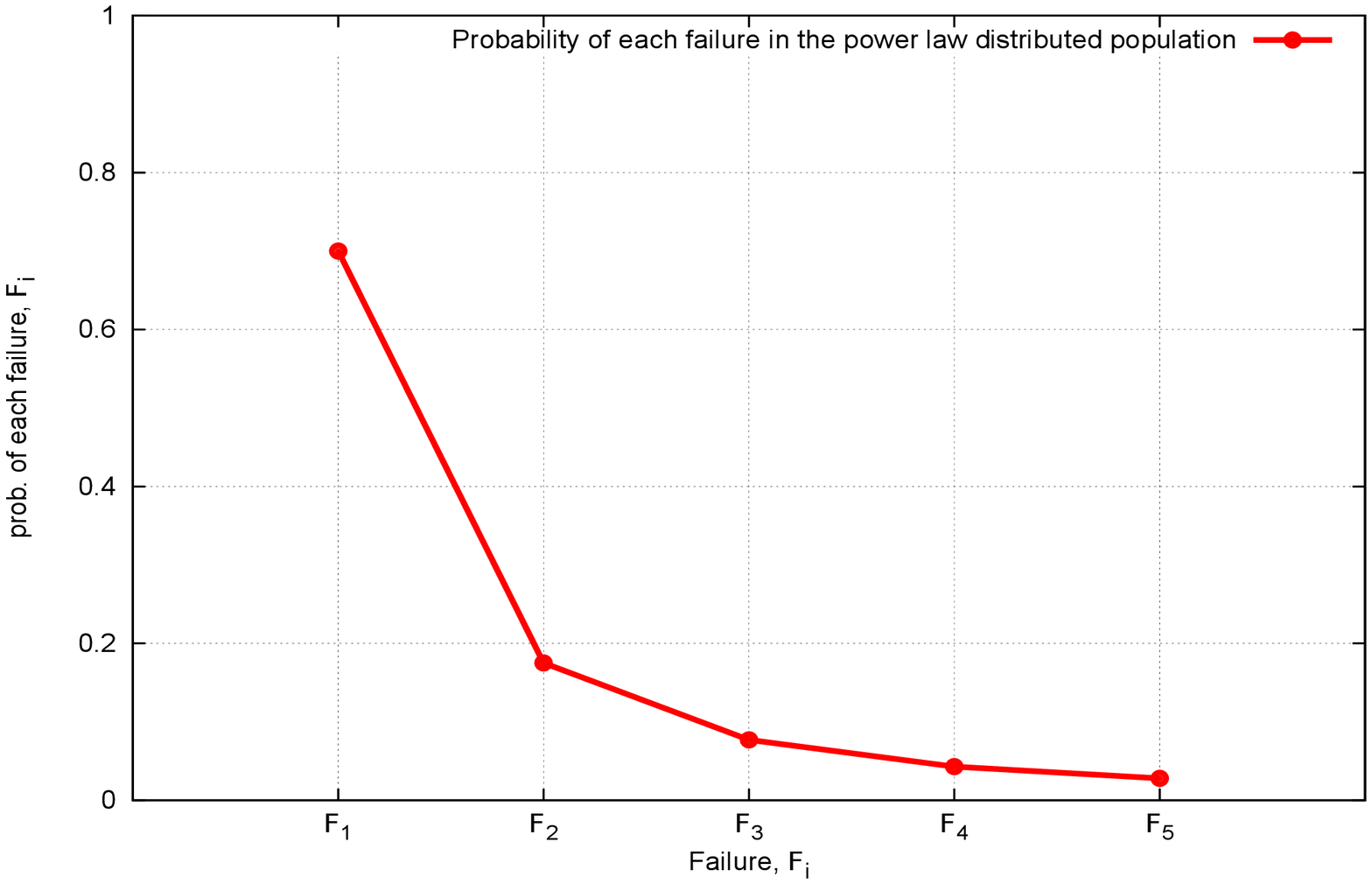}
\caption{Probability of each failure $F_i$ in the power law distributed population}
\label{fig_plot_prob_of_each_failure_inkscape}
\end{figure}

\begin{figure}[ht]
\centering
\includegraphics[width=\columnwidth]{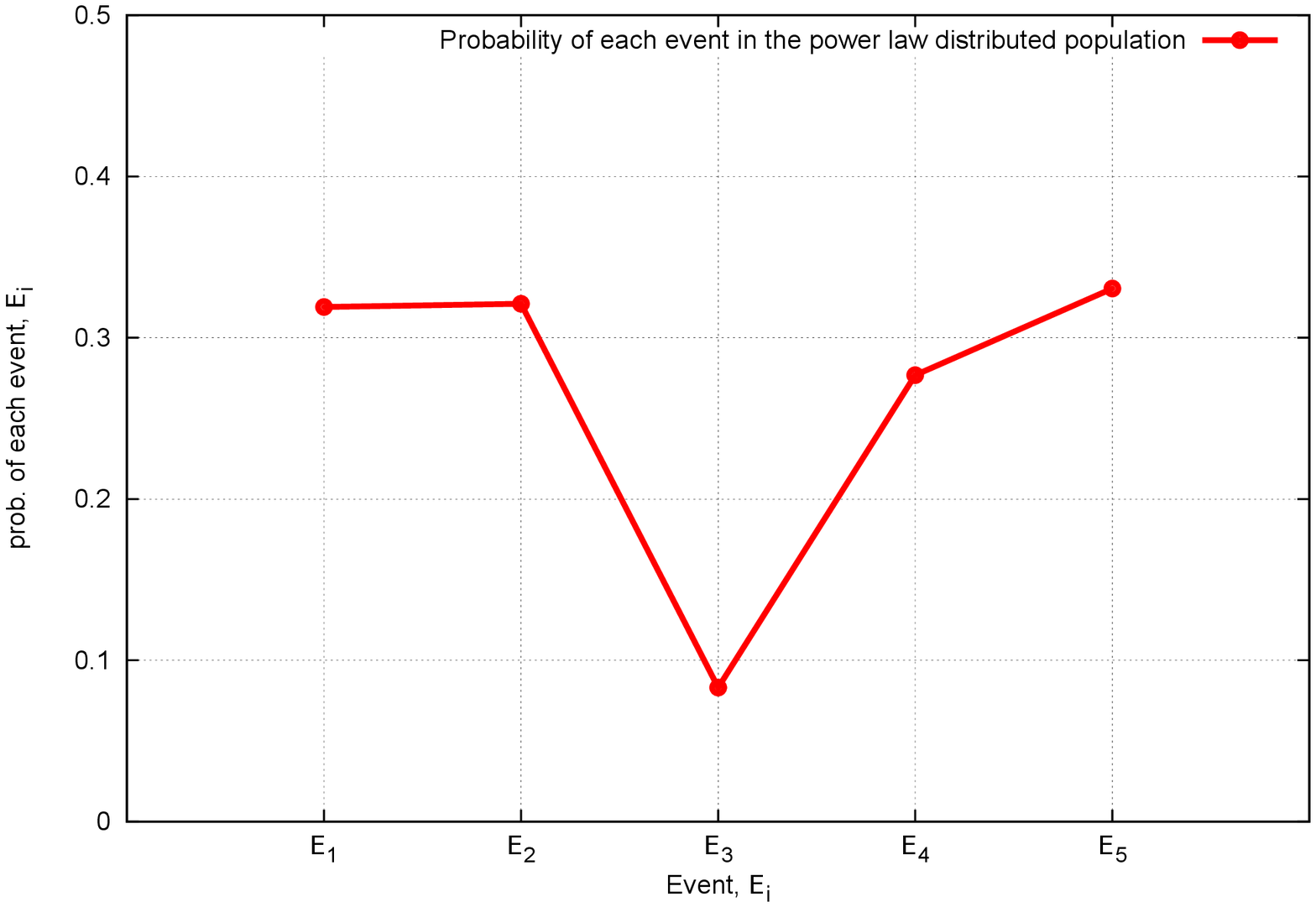}
\caption{Probability of each event $E_i$ in the power law distributed population}
\label{fig_plot_prob_of_each_event_inkscape}
\end{figure}

For generation of population of failures the power law distribution in (\ref{eqn_power_law}) is used with $k = 2$, $N = 5$, $a = 0.7$ satisfying (\ref{eqn_total_prob}). Number of each failures $F_i$, $i = 1, 2, 3, 4, 5$ is shown in Fig. \ref{fig_plot_number_of_each_event_inkscape} and the probability distribution of failures is shown in Fig. \ref{fig_plot_prob_of_each_failure_inkscape}, under the assumption that occurrence frequency of $F_1$ $>$ occurrence frequency of $F_2$ $>$ occurrence frequency of $F_3$ $>$ occurrence frequency of $F_4$ $>$ occurrence frequency of $F_5$, available from the qualitative information provided by the developers. Ten thousand samples of failures are generated. The probabilities of the events are shown in Fig. \ref{fig_plot_prob_of_each_event_inkscape}.
\subsection{Application of BN}
Using the augmented data described above in section \ref{section_results_generate_samples}, the conditional probabilities necessary for prediction of failures applying BN are calculated. Probabilities of $E_1$ are shown in Table \ref{table_pr_E_1}.
\begin{table}[ht]
  \caption{Probabilities of $E_1$}
  \centering
  \begin{tabular}{|c|c|}
  \hline
  Pr($E_1$ = 1) & Pr($E_1$ = 0) \\  [0.5ex]
  \hline
  0.924128 & 0.075871 \\
  \hline
  \end{tabular}
  \label{table_pr_E_1}
\end{table}
Similarly, the probabilities of $E_2$ given its parent $E_1$ (Fig. \ref{fig_DAG_for_BN_inkscape}) occurred or not are provided in Table \ref{table_pr_E_2}. Likewise, the same for $E_3$ is provided in Table. \ref{table_pr_E_3}. Note that probabilities of $E_3$ given $E_1 = 0$ and $E_2 = 0$ are not valid failures in current set (\ref{eqn_event_failure_table}). However, \verb|pgmpy| needs all the combinations of probabilities of nodes given their parents to be made available and each row in the tables should add up to 1. This does not adversely affect the performance of the prediction model as the results show subsequently. The probabilities of $E_4$ and $E_5$ given their respective parents are shown in Tables \ref{table_pr_E_4} and \ref{table_pr_E_5}. These probabilities are then fed into the BN for failure prediction.
\begin{table}[ht]
  \caption{Probabilities of $E_2$}
  \centering
  \begin{tabular}{|c|c|c|c|}
  \hline
  Condition on & Pr($E_2$ = 1) & Pr($E_2$ = 0) & Comments\\  [0.5ex]
  \hline
  $E_1$ = 1 & 0.924128 & 0.075871 & \\
  \hline
  $E_1$ = 0 & 1 & 0 & \vtop{\hbox{\strut Tool needs all}\hbox{\strut the combination}}\\
  \hline
  \end{tabular}
  \label{table_pr_E_2}
\end{table}

\begin{table}[ht]
  \caption{Probabilities of $E_3$}
  \centering
  \begin{tabular}{|c|c|c|c|}
  \hline
  Condition on & Pr($E_3$ = 1) & Pr($E_3$ = 0) & Comments\\  [0.5ex]
  \hline
  $E_1$ = 0 and $E_2$ = 0 & 0 & 1 & \vtop{\hbox{\strut Tool needs all}\hbox{\strut the combinations}\hbox{\strut to add up to 1}}\\
  \hline
  $E_1$ = 0 and $E_2$ = 1  & 0 & 1 & \\
  \hline
  $E_1$ = 1 and $E_2$ = 0  & 1 & 0 & \\
  \hline
  $E_1$ = 1 and $E_2$ = 1  & 0.2 & 0.8 & \\
  \hline
  \end{tabular}
  \label{table_pr_E_3}
\end{table}

\begin{table}[ht]
  \caption{Probabilities of $E_4$}
  \centering
  \begin{tabular}{|c|c|c|c|}
  \hline
  Condition on & Pr($E_4$ = 1) & Pr($E_4$ = 0) & Comments\\  [0.5ex]
  \hline
  $E_2$ = 0 and $E_3$ = 0 & 0 & 1 & \vtop{\hbox{\strut Tool needs all}\hbox{\strut the combinations}\hbox{\strut to add up to 1}}\\
  \hline
  $E_2$ = 0 and $E_3$ = 1  & 0.607843 & 0.392156 & \\
  \hline
  $E_2$ = 1 and $E_3$ = 0  & 1 & 0 & \\
  \hline
  $E_2$ = 1 and $E_3$ = 1  & 0 & 1 & \\
  \hline
  \end{tabular}
  \label{table_pr_E_4}
\end{table}

\begin{table}[ht]
  \caption{Probabilities of $E_5$}
  \centering
  \begin{tabular}{|c|c|c|c|}
  \hline
  Condition on & Pr($E_5$ = 1) & Pr($E_5$ = 0) & Comments\\  [0.5ex]
  \hline
  $E_3$ = 0 and $E_4$ = 0 & 0 & 1 & \vtop{\hbox{\strut Tool needs all}\hbox{\strut the combinations}\hbox{\strut to add up to 1}}\\
  \hline
  $E_3$ = 0 and $E_4$ = 1  & 1 & 0 & \\
  \hline
  $E_3$ = 1 and $E_4$ = 0  & 1 & 0 & \\
  \hline
  $E_3$ = 1 and $E_4$ = 1  & 0 & 1 & \\
  \hline
  \end{tabular}
  \label{table_pr_E_5}
\end{table}

Non-occurrence of an event, i.e., $E_1$ = 0, is hard to provide as evidence to the BN. Hence, occurrence of event is always set as evidence to predict the failures. The tool takes all the probabilities provided in that tables above and outputs the prediction after calculating the equations such as (\ref{eqn_example_predict_3}).

If function call to query the BN for prediction of the subsequent events with the evidence that $E_1$ has already occurred, its output predicts $F_1$ defined in (\ref{eqn_event_failure_table}) as shown below. The \verb|PREDICTION| is  concatenation of \verb|evidence| and \verb|OUTPUT|.

\begin{Verbatim}[tabsize=4]
FUNCTION CALL:
    infer.map_query(['E2', 'E3',
                     'E4', 'E5'],
                    evidence={'E1': '1'})
OUTPUT:
    {'E2': '1', 'E3': '0',
     'E4': '1', 'E5': '1'}
PREDICTION:
    {'E1': '1', 'E2': '1', 'E3': '0',
     'E4': '1', 'E5': '1'} --> Failure F1
\end{Verbatim}
Afterwards, when events $E_1$ and $E_2$ occur which are presented as evidence to the BN, it continues to predict failure $F_1$.
\begin{Verbatim}[tabsize=4]
FUNCTION CALL:
infer.map_query(['E3', 'E4', 'E5'],
                evidence={
                'E1': '1',
                'E2': '1'})
OUTPUT:
    {'E3': '0', 'E4': '1', 'E5': '1'}
PREDICTION:
    {'E1': '1', 'E2': '1', 'E3': '0',
     'E4': '1', 'E5': '1'} --> Failure F1
\end{Verbatim}
With evidence $E_1$, $E_2$ and $E_3$, the BN changes its prediction from  $F_1$ to $F_2$ as defined in (\ref{eqn_event_failure_table}).
\begin{Verbatim}[tabsize=4]
FUNCTION CALL:
infer.map_query(['E4', 'E5'],
                evidence={
                'E1': '1',
                'E2': '1',
                'E3': '1'})
OUTPUT:
    {'E4': '0', 'E5': '1'}
PREDICTION:
    {'E1': '1', 'E2': '1', 'E3': '1',
     'E4': '0', 'E5': '1'} --> Failure F2
\end{Verbatim}
However, if  occurrence of $E_1$, $E_2$, $E_3$ and $E_4$ are provided as evidence then it correctly detects an invalid event since the output does not match with any row in (\ref{eqn_event_failure_table}).
\begin{Verbatim}[tabsize=4]
FUNCTION CALL:
infer.map_query(['E5'],
                evidence={
                'E1': '1',
                'E2': '1',
                'E3': '1',
                'E4': '1'})
OUTPUT:
    {'E5': '0'}
PREDICTION:
    {'E1': '1', 'E2': '1',
     'E3': '1', 'E4': '1',
     'E5': '0'} --> invalid event
\end{Verbatim}
If occurrence of events $E_2$ and $E_4$ are provided as evidence then the BN predicts $F_3$ as expected.
\begin{Verbatim}[tabsize=4]
FUNCTION CALL:
infer.map_query(['E5'],
                evidence={
                'E2': '1',
                'E4': '1'})
OUTPUT:
    {'E5': '1'}
PREDICTION:
    {'E1': '0', 'E2': '1', 'E3': '0',
     'E4': '1', 'E5': '1'} --> Failure F3
\end{Verbatim}
If events $E_1$ and $E_3$ are evidences, then $F_4$ is predicted due to its higher probability (Fig. \ref{fig_plot_prob_of_each_failure_inkscape}).
\begin{Verbatim}[tabsize=4]
FUNCTION CALL:
infer.map_query(['E4', 'E5'],
                evidence={
                'E1': '1',
                'E3': '1'})
OUTPUT:
    {'E4': '1', 'E5': '0'}
PREDICTION:
    {'E1': '1', 'E2': '0', 'E3': '1',
     'E4': '1', 'E5': '0'} --> Failure F4
\end{Verbatim}
To predict $F_5$, the query has to happen on the evidence that $E4$ has occurred, since it is the only difference between $F_4$ and $F_5$, and $E_1$, $E_3$ and $E_5$ have to occur. Doing so, the BN predicts $F_5$ correctly.
\begin{Verbatim}[tabsize=4]
FUNCTION CALL:
infer.map_query(['E4'],
                evidence={
                'E1': '1',
                'E3': '1',
                'E5': '1'})
OUTPUT:
    {'E4': '0'}
PREDICTION:
    {'E1': '1', 'E2': '0', 'E3': '1',
     'E4': '0', 'E5': '1'} --> Failure F5
\end{Verbatim}
\section{Conclusion and Future Work}\label{section_conclusion}
Failures in backbone optical networks can lead to major disruption in internet traffic. Hence, prediction of such failures can avoid such problems. This paper proposed an data augmented BN to predict failures of networks node using some information from logs and (qualitative) inputs from developers on frequency of occurrence of failures. The conditional probabilities of the BN is calculated after generation of failure population applying a power law distribution of the failures based on their frequency of occurrences. Results show that the proposed node failure prediction mechanism is able to perform with high accuracy.

Future work will extend the model to more nodes in the BN and integrate this to the deployed network.

\section*{Acknowledgment}
This research project is funded by Tejas Networks, Bangalore, India.

\bibliographystyle{IEEEtran}
\bibliography{IEEEabrv,net_failure_pred}
\end{document}